\documentclass[a4paper,11pt]{article}
\pdfoutput=1 % if your are submitting a pdflatex (i.e. if you have
\usepackage{jinstpub} % for details on the use of the package, please
\title{\boldmath The SHiP experiment and the RPC technology}

\author[a,b,c]{G. De Lellis,\note{on behalf of the SHiP Collaboration}}
\affiliation[a]{University "Federico II" of Naples,\\via Cintia, Naples, Italy}
\affiliation[b]{INFN,\\via Cintia, Naples, Italy}
\affiliation[c]{MISiS,\\Leninsky Prospect 4, Moscow}

\emailAdd{giovanni.de.lellis@cern.ch}

\abstract{The discovery of the Higgs boson has fully confirmed the Standard Model of particles and fields. Nevertheless, there are still fundamental phenomena, like the existence of dark matter and the baryon asymmetry of the Universe, which deserve an explanation that could come from the discovery of new particles. The SHiP experiment at CERN meant to search for very weakly coupled particles in the few GeV mass domain has been recently proposed. The existence of such particles, foreseen in Beyond Standard Models, is largely unexplored. A beam dump facility using high intensity 400 GeV protons is a copious source of such unknown particles in the GeV mass range. The beam dump is also a copious source of neutrinos and in particular it is an ideal source of tau neutrinos, the less known particle in the Standard Model. We report the physics potential of such an experiment and describe the use of the RPC technology therein. An anchillary measurement of the charm cross-section will be carried out in July 2018 and RPC are used as a muon detector. We also describe the design and construction of these new chambers.
}


\arxivnumber{1234.56789} % only if you have one

\proceeding{14$^{\text{th}}$ Workshop on Resistive Plate Chambers and related detectors\\
  19-23 February 2018\\
  Puerto Vallarta, Jalisco State, Mexico}

\begin{document}
\maketitle
\flushbottom

\section{The SHiP experiment}
\label{sec:ship}

The discovery of the Higgs boson is certainly a big triumph of the Standard Model. In particular, given its mass, it could well be that the Standard Model is an effective theory working all the way up to the Planck scale. Nevertheless, there are several phenomena deserving an explanation that the Standard Model is unable to provide: the existence of dark matter and its nature, the baryonic asymmetry of the Universe and neutrino masses. It is therefore clear the new physics is there and presumably several new particles have still to be discovered.

Searches for new physics with accelerators are being carried out at the LHC, especially suited to look for high mass particles with ordinary couplings to matter. Complementary searches for very weakly coupled and therefore long-lived particles require a beam dump facility. Such a facility is made of a high density proton target, followed by a hadron stopper and a muon shield. Apart from residual muons, the only remaining particles are electron, muon and tau neutrinos on top of hidden, long-lived particles produced either in proton interactions or in secondary particle decays.

A new experiment, Search for Hidden Particles (SHiP), has been proposed~\cite{ship_tp}, designed to operate at a beam dump facility to be built at CERN and to search for weakly coupled particles in the few GeV mass range. The physics case for such an experiment is widely discussed
in Ref.~\cite{ship_pp}. In five years, the facility will integrate $2 \times 10^{20}$ 400 GeV protons, produced by the SPS accelerator complex, impinging on a 12 interaction length ($\lambda_{int}$) target made of Molybdenum and Tungsten, followed by a 30 $\lambda_{int}$ iron hadron absorber. Hidden particles in the GeV mass range would be produced mostly by the decay of charmed hadrons produced in proton interactions. $D_s$ mesons, copiously produced among charmed hadrons, are a source of tau neutrinos through their fully leptonic decay. Therefore, the SHiP facility is ideal also to study the physics of tau neutrinos, the less known particle in the Standard Model.

\begin{figure}[h]
% Use the relevant command for your figure-insertion program
% to insert the figure file.
\centering
\includegraphics[width=0.95\textwidth]{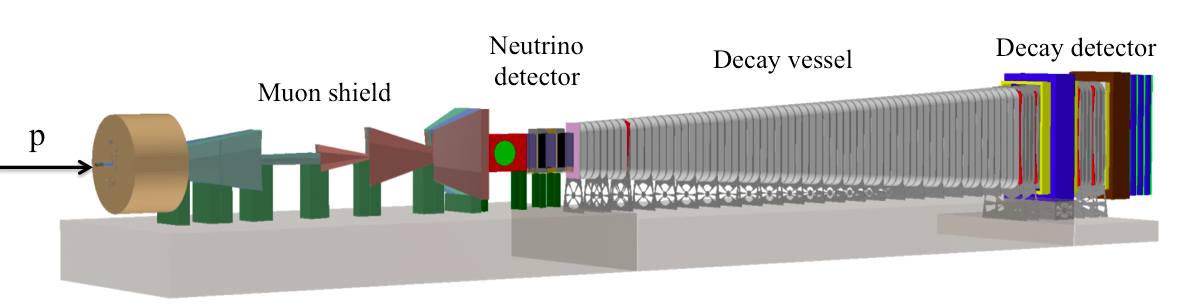}
\caption{The beam dump facility and the SHiP detector.
}
\label{fig-1}       % Give a unique label
\end{figure}

Figure~\ref{fig-1} shows the SHiP facility to be placed in the North Area: downstream of the target, the hadron absorber filters out all hadrons, therefore only muons and neutrinos are left. An active muon shield is designed with two sections with opposite polarity to maximize the muon flux reduction: it reduces the muon flux from $~10^{10}$ down to $~10^5$ muons per spill. $4 \times 10^{13}$ protons are extracted in each spill, designed to be 1s long to reduce the detector occupancy~\cite{muon_shield}. A first successful test of the SPS cycle with a 1s long spill was performed in April 2015.

The neutrino detector is located downstream of the muon shield, followed by the decay vessel and the detector for hidden particles. The Collaboration will prepare a document for the European Strategy by the end of 2018 and a Comprehensive Design Report by 2019, in the framework of the Physics Beyond Colliders working group, launched in 2016 at CERN. The construction and installation of the detector will start in 2021 and last until the end of the third LHC long shutdown such that the data taking is assumed to start in 2026.

\begin{figure}[h]
% Use the relevant command for your figure-insertion program
% to insert the figure file.
\centering
\includegraphics[width=0.9\textwidth]{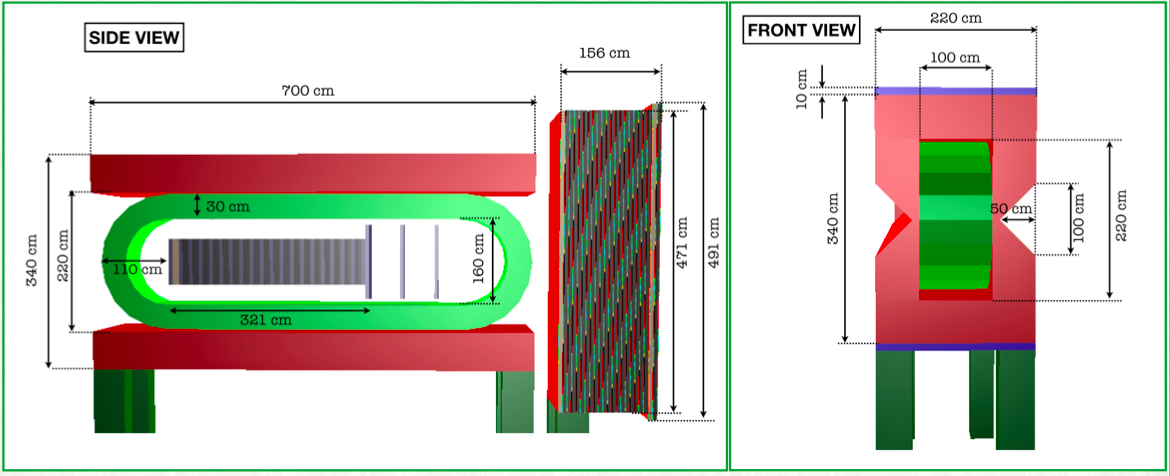}
\caption{The neutrino detector upstream of the decay vessel in different views.
}
\label{fig-2}       % Give a unique label
\end{figure}

The neutrino detector is made of a magnetised  region, followed by a muon identification system, as shown in Figure~\ref{fig-2}. The magnetised region will host both the neutrino target and a particle spectrometer. The neutrino target is based on the emulsion cloud chamber technology employed by the OPERA experiment [4], with a compact emulsion spectrometer, made of a sequence of very low density material and emulsion films to measure the charge and momentum of hadrons in magnetic field. This detector is suitable for the measurement of momenta up to 12 GeV$/c$. Indeed, this feature would allow to discriminate between tau neutrinos and anti-neutrinos also in the hadronic decay channels of the tau lepton. The emulsion target is complemented by high resolution tracking chambers to provide the time stamp to the event, connect muon tracks from the target to the muon system and measure the charge and momentum for particles with momenta above 10 GeV$/c$. The muon system is based on 23 iron slabs, 5 cm thick each, alternated with 24 RPCs  providing the tracking within the slabs. The muon system will also act as upstream veto tagger for background processes to the hidden particle search, which motivates the high sampling choice. Nevertheless, the muon system configuration is still under optimisation.

The emulsion target will also act as the target of dark matter as well as of any very weakly in- teracting particle produced at the accelerator, when its mass is in the GeV range.  The ongoing optimisation of this detector concerns the target material, the sampling frequency of the emulsion cloud chamber and the timing performances of the target tracker that would enable the separation between neutrinos and heavy particles based on the time of flight measurement.

The detector for hidden particles is located in the downstream part of a 60 m long evacuated decay vessel, with a conical shape and an elliptical transverse section at the very downstream end of $5 \times 10$~m$^2$, the longer axis being vertical. The hidden particles are supposed to decay within the vessel. The requirement to have less than 1 neutrino interaction in the vessel over five years sets the pressure to about $10^{-3}$ mbar. A magnetic spectrometer is located downstream of the vessel: it is made of straw tubes with a material budget of 0.5\% $X_0$ per station, achieving a position resolution of 120 $\mu$m per straw, with 8 hits per station on average. This gives a momentum resolution of about 1\%. The vessel would be sorrounded by a liquid scintillator layer to tag particles coming from outside. Downstream of the spectrometer, an hadronic and electromagnetic calorimeter and a muon filter are used to identify particles. A timing detector complements the apparatus to reject vertices from chance coincidences. 

\section{Search for hidden particles and physics with the neutrino detector}
\begin{figure}[h]
% Use the relevant command for your figure-insertion program
% to insert the figure file.
\centering
\includegraphics[width=0.52\textwidth]{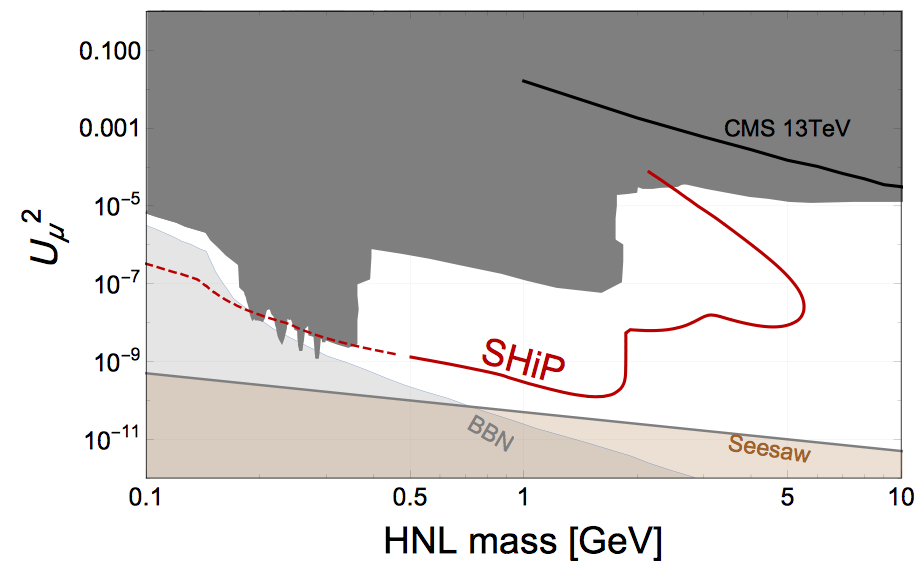}
\includegraphics[width=0.47\textwidth]{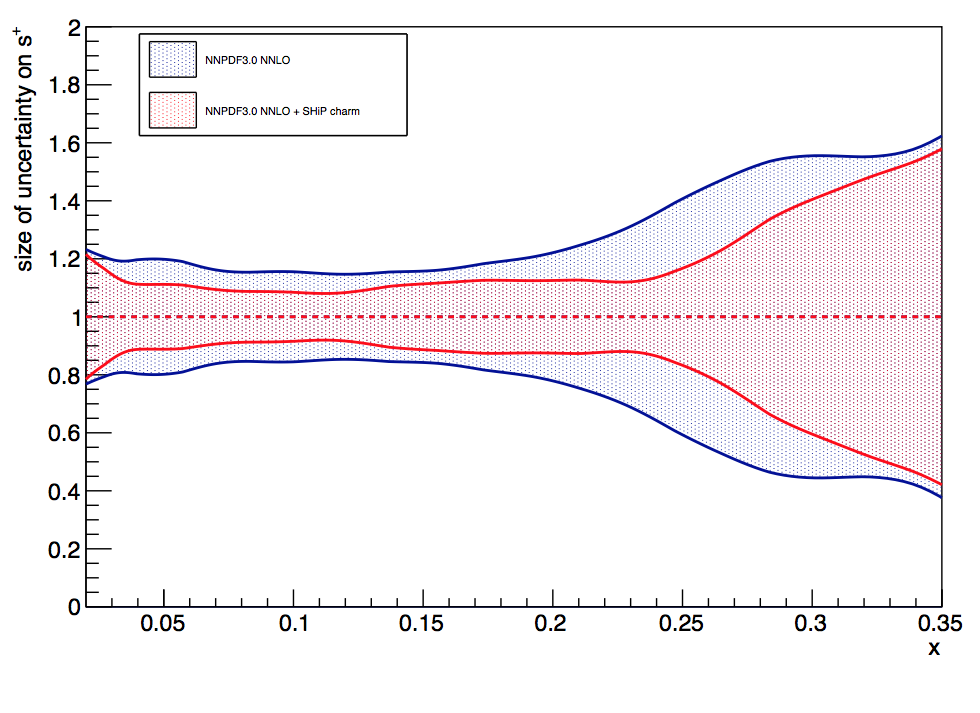}
\caption{Left: SHiP sensitivity to heavy neutral leptons~\cite{oleg}. Right: Improvement of the accuracy on $s^+$ with SHiP (red) compared to the present status (blue) in the $0.02 < x < 0.35$ range.
}
\label{fig-3}       % Give a unique label
\end{figure}

Extensions of the Standard Model in the low mass region foresee the existence of particles as singlets with respect to the Standard Model gauge group. These particles couple to different singlet composite operators (so-called Portals) of the Standard Model. The SHiP detector has the potentiality to discover very weakly interacting and long lived particles in a wide unexplored range of their masses and couplings, within these Portals. As an example, we report in the left plot of Figure~\ref{fig-3} the sensitivity to heavy neutral leptons, when only the muon coupling  $U_\mu$ is considered~\cite{oleg}. For an overview of the sensitivity to different portals and corresponding particles, we refer to~\cite{ship_tp,ship_pp}.

The observation of tau neutrinos was  confirmed  by the DONUT experiment only in 2008 when 9 candidates events were reported~\cite{donut2}.  The OPERA experiment~\cite{opera} has detected ten tau neutrinos~\cite{nutau1,nutau2,nutau3,nutau4,nutau5,finaltau}, leading to the discovery of tau neutrino appearance from muon neutrino oscillations~\cite{nutau5,finaltau}. The only leptonic decay observed by OPERA~\cite{nutau3} shows negative charge as expected from a $\nu_{\tau}$ interaction. Therefore, so far there is no direct evidence for tau anti-neutrinos.
The SHiP facility is  a $\nu_{\tau}$ factory, with $6.6 \times 10^{15}$ tau neutrinos produced in primary proton collisions, equally divided in neutrinos and anti-neutrinos. Given the neutrino target mass of about 10 tons, one expects more than 10000 interactions of tau neutrinos and anti-neutrinos.

Charmed hadrons are produced in neutrino and anti-neutrino charged-current interactions at the level of about 5\%~\cite{charm}. Experiments based on calorimetric technology identify charmed hadrons only in their muonic decay channel, when two opposite sign muons are produced in the final state. A cut of 5~GeV is applied to muons in order to suppress the background due to punch-through pions. The nuclear emulsion technology, instead, identifies topologically the charmed hadron by detecting its decay vertex. Energy cuts are therefore much looser, thus providing a better sensitivity to the charm quark mass. Moroever, a large statistical gain is provided by the use of hadronic decay modes~\cite{charm}. Indeed,  SHiP will integrate about $10^5$ charm candidates, more than one order of magnitude larger than the present statistics, with a large ($\sim 30$\%) contribution from anti-neutrinos. Charm production in neutrino scattering is extremely sensitive to the strange quark content of the nucleon, especially with anti-neutrinos where the $s$-quark is dominant. SHiP will improve significantly the uncertainty on the strange quark distribution in the nucleon as shown in the right plot of Figure~\ref{fig-3} in terms of $s^+ = s(x) + \bar{s}(x)$ in the $0.02 < x < 0.35$ range.

\section{RPC technology in SHiP}
The RPC technology is proposed to be used for two different applications in SHiP: one is a tracking detector for the muon system of the neutrino detector, also acting as a veto for the background of hidden particle decays. Prototypes of these chambers are being constructed for the measurement of the charm cross-section, where RPC will instrument the muon system. The other application of the RPC technology in SHiP is the timing detector in the downstream apparatus for the detection of hidden particles. We describe both applications here. 

\begin{figure}
\centering
\includegraphics[width=.7\textwidth,origin=c]{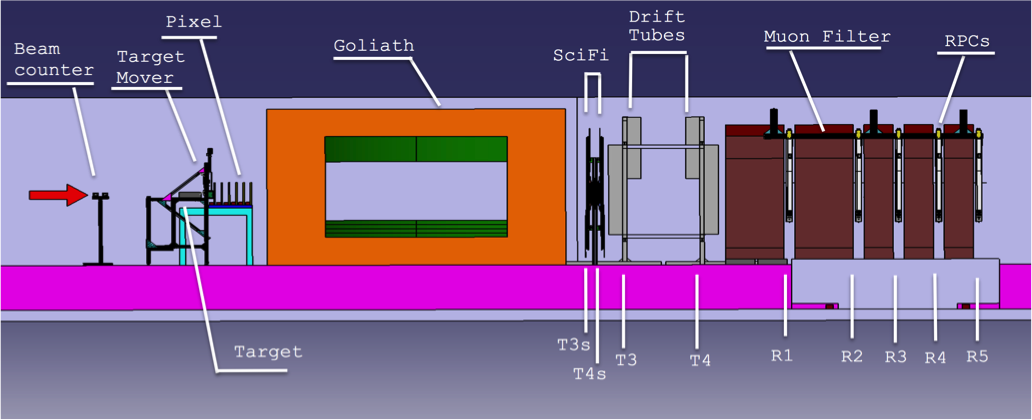}
\includegraphics[width=.29\textwidth,origin=c]{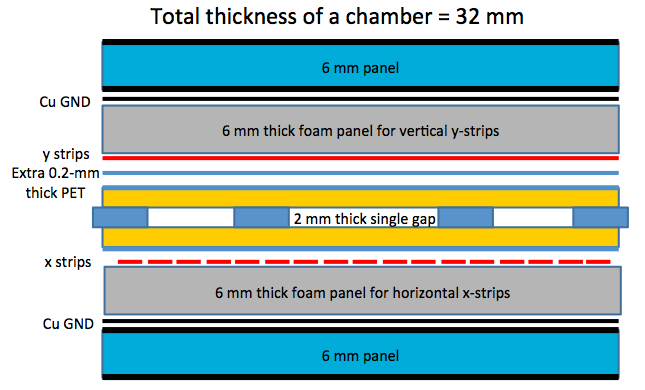}
\caption{Left:
Setup of the charm measurement experiment including the downstream muon filter based on the RPC technology. Right: structure of the 32 mm thick RPC chamber.}
\label{fig5}       % Give a unique label
\end{figure} 

\subsection{Muon system for the charm cross-section measurement}
The prediction of the tau neutrino yield is affected by a large uncertainty: indeed, simulation studies of proton interactions in heavy and thick targets show that the charmed hadron yield is increased by a factor of 2.3 from the cascade production~\cite{ship_note}. Charmed hadrons are produced either directly from interactions of the primary protons or from subsequent interactions of the particles produced in the hadronic cascade showers, including the protons after a primary elastic collision. The only available measurement of the associated charm production  per nucleon $\sigma_{c\bar{c}} = 18.1 \pm1.7$ $\mu$barn~\cite{doublecharm} was indeed obtained with a thin target where the secondary production is negligible. 

The SHiP Collaboration has proposed the SHiP-charm project~\cite{ship_charm_eoi}, aiming at measuring the associated charm production by employing the SPS 400 GeV/c proton beam. This proposal includes a study of the cascade effect to be carried out using ECC techniques, i.e.~slabs consisting of a replica of the SHiP experiment target~\cite{ship_tp} interleaved with emulsion films. The detector is hybrid, combining the emulsion technique with electronic detectors to provide the charge and momentum measurement of charmed hadron decay daughters and the muon identification. This setup shown in the left part of Figure~\ref{fig5} allows a full kinematical reconstruction of the event.  An optimisation run was approved at CERN for July 2018 while the full measurement is planned after the long shutdown LS2 of the CERN accelerator complex, with $5 \times 10^7$ protons on target and a charm yield of about 2500 fully reconstructed interactions.

The RPC chambers were designed to operate in avalanche mode, with a time resolution of about 1 ns. Two orthogonal sets of strips are used for 2D  measurements with an expected position resolution of about 3~mm in both directions. Their structure is shown in the right part of Figure~\ref{fig5}. The 2D measurement is necessary to cope with the large occupancy of the detector: indeed, in each event there are on average several tens of particles entering the first two RPC detectors of the muon system, only one or two being a muon particle. The bakelite electrodes were produced at Puricelli s.r.l.~in Italy. Each RPC is using 118 horizontal and 184 vertical strips with a pitch of 10.625 mm  produced at KODEL in Korea, with a total size of $2100 \times 1350$ mm$^2$ and an active area of $1900 \times 1200$~mm$^2$. Each RPC is equipped with 20 front-end cards, FEERIC developed by the ALICE Collaboration, 12 connected to vertical and 8 to horizontal strips. 5 readout boards are used for each chamber. In total 5 chambers were built and are being tested at CERN before their installation at the H4 beamline at CERN in July 2018.

\subsection{Timing detector}
One of the SHIP timing detector prototypes is based on timing Resistive Plate Chamber (tRPC) technology. The prototype uses a novel concept, i.e.~the RPC sensitive volume. With this approach, the gas volume and the High Voltage (HV) insulation are confined inside a permanent sealed plastic box, decoupling it from the pick up electrodes located on the top and on the bottom of the sensitive volume. The main advantages of this sensitive volume are: versatility, the same volume can be coupled to different readout electrodes; ease of construction and low cost on the $1 \div 2$~m$^2$ scale with a complete tightness of the plastic box allowing an operation with low gas flux.

The sensitive volume of the SHIP tRPC prototype houses a multi-gap RPC structure`\cite{mrpc} with six gas gaps defined by seven 1 mm thick float glass  electrodes of about $1550 \times 1250$~mm$^2$ separated by 0.3 mm nylon mono-filaments. The HV electrodes are made up of a resistive layer  applied to the surface of the outermost glasses with airbrush techniques. The structure is permanently sealed inside a PMMA gas tight box with a 1 mm lid thickness equipped with feed-throughs for gas and HV connections. 

The RPC chamber is composed of two identical sensitive modules, read out by a pick-up electrode, located between the modules, made from FR4 Printed Circuit Board with a thickness of 1.5 mm and equipped with $1600 \times 30$~mm$^2$ copper strips. The set is enclosed in an aluminium case to guarantee the electromagnetic insulation from the environment and enough mechanical rigidity. The chamber was operated with pure C$_2$H$_2$F$_4$ and open gas flow.
Both sides of each strip are directly connected to low-jitter high-gain/bandwidth Front-End Electronics~\cite{fee} and its digital output connected to a FPGA based multi-hit TDC~\cite{tdc}. The time of the each strip is calculated as the average of the times in each side.

\begin{figure}
\centering
\includegraphics[width=.9\textwidth,origin=c]{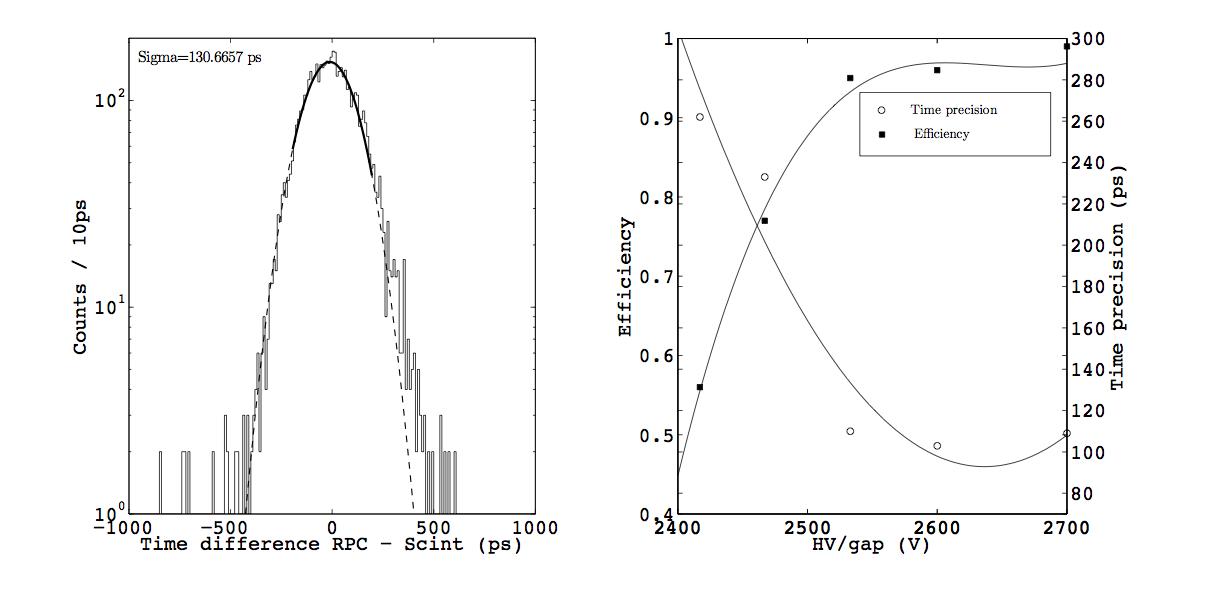}
\caption{
Left: Distribution of the time difference between one of the reference scintillators and the prototype chamber after the walk correction. Right: Time accuracy and efficiency as a function of the HV/gap.}
\label{fig-4}       % Give a unique label
\end{figure}

Three fast plastic scintillators (BC420, $2 \times 3 \times 8$~cm$^3$) readout on both sides by Hamamatsu H6533 photomultipliers are used to trigger on cosmic muons and to provide a time reference, in order to evaluate the response of the prototype. The three scintillators are aligned with one of the strips, two above and one below the chamber. The left plot of Figure~\ref{fig-4} shows the time distribution of the difference between one of the reference scintillators and the prototype after the walk correction. The time precision of the chamber after subtracting the contribution of the scintillator ($\sigma \sim 107$~ps)  shows a $\sigma \sim 105$~ps . The right plot of Figure~\ref{fig-4} shows the time accuracy and the chamber efficiency as a function of the HV/gap:  a plateau in the efficiency is reached above 2550 V/gap as well as an accuracy of about 105 ps. At the plateau, the dark counting rate of the detector is about 2.2~kHz/m$^2$. These measurements were carried out 
at the University of Coimbra in Portugal and further tests will be performed with the final gas mixture 90\%  C$_2$H$_2$F$_4$ and 10\% SF6. Nevertheless, the performance already measured makes the tRPC a good candidate to instrument the 50 m$^2$ SHIP timing detector.

% We suggest to always provide author, title and journal data:
% in short all the informations that clearly identify a document.

\end{document}